\documentclass[12pt]{article}  
\usepackage{cite}
\def \bea{\begin{eqnarray}}
\def \beq{\begin{equation}}
\def \eea{\end{eqnarray}}
\def \eeq{\end{equation}}
\def \esix{E$_{\rm 6}$}

\def \s{\sqrt{2}}

\begin{document}

\rightline{EFI 14-8}
\rightline{arXiv:1404.5198}
\centerline{\bf THREE STERILE NEUTRINOS IN E$_6$}
\medskip
\centerline{\it Jonathan L. Rosner}
\medskip
\centerline{\it Enrico Fermi Institute and Department of Physics}
\centerline{\it University of Chicago, Chicago, IL 60637}
\bigskip

\begin{quote}

Candidates for unfication of the electroweak and strong interactions include
the grand unified groups SU(5), SO(10), and the exceptional group \esix.  The
27-dimensional fundamental representation of \esix~contains exotic fermions,
including weak isosinglet quarks of charge --1/3, vector-like weak
isodoublet leptons, and neutral leptons which are singlets under
both left-handed and right-handed SU(2).  These last are candidates for
light ``sterile'' neutrinos, hinted at by some recent short-baseline neutrino
experiments.  In order to accommodate three families of quarks and charged
leptons, an \esix~model must contain three 27-plets, each of which contains a
sterile neutrino candidate $n$.  The mixing pattern within a 27-plet is
described, and experimental consequences are discussed.

\end{quote}

\leftline{PACS categories: 12.10.Dm, 12.60.Cn, 14.60.Pq, 14.60.St}
\bigskip

\centerline{\bf I.  INTRODUCTION}
\bigskip

The standard model group SU(3)$_{\rm color} \times$ SU(2)$_{\rm L} \times$ U(1)
can be incorporated into a grand unified group.  Candidates include SU(5),
SO(10), and \esix~\cite{GRS}.  Each quark and lepton family consists of a
$5^* + 10$ representation of SU(5).  Adding a right-handed neutrino $N$ [an
SU(5) singlet] to each such hypermultiplet, one gets a spinor 16-plet of
SO(10).  A right-handed neutrino can pair with a left-handed one to
generate a Dirac mass $m_D$ as occurs for charged leptons and quarks.  However,
the neutrality of the right-handed neutrino under the standard model group
allows it to have a large Majorana mass $M$, leading via the seesaw mechanism
\cite{seesaw} to light-neutrino masses $m_\nu = m_D^2/M$.  At this stage there
are three light neutrinos (mostly electroweak doublets) and three heavy ones
(mostly electroweak singlets).

In addition to the ``active'' light neutrinos $\nu_e,\nu_\mu,\nu_\tau$, some
short-baseline neutrino experiments \cite{sbl1,sbl2,sbl3,Kopp:2013vaa,%
deGouvea:2013onf,Aguilar:2001ty,AguilarArevalo:2007it,Mention:2011rk,
Mueller:2011nm,Huber:2011wv} have hinted at the existence of one or more light
``sterile'' neutrinos, participating in the weak interactions only via mixing
with the active ones.  For schemes in which one or more of the right-hand
neutrinos plays the role of a light sterile neutrino, see
\cite{Boyarsky:2009ix,Merle:2013gea}.  These typically involve constraints
in comparison with scenarios in which all three right-handed neutrinos are
heavy.

In this paper we wish to investigate a different scenario for sterile neutrinos,
based on \esix.  This would be especially timely if the exotic states predicted
in \esix~were to show up in forthcoming experiments at the CERN Large Hadron
Collider.  A 10-plet of SO(10) [a 5 + $5^*$ of SU(5)] can be added to
each quark and lepton family.  It consists of quarks $h + h^c$ which are
singlets under SU(2)$_L$ and SU(2)$_R$, and leptons $E^\pm$ and their neutral
counterparts which are doublets under both.  The smallest
\esix~representation, a 27-plet, is formed by adding another singlet $n$ of
SO(10).  The $n$ has neither L nor R isospin.  We shall explore a scenario in
which the three $n$ states are candidates for light sterile neutrinos, leaving
all three right-handed neutrinos unconstrained and potentially very heavy.

Fits to short-baseline neutrino anomalies include ones in Refs.\ \cite{sbl2,%
sbl3,Kopp:2013vaa}.  There is general agreement that at least two sterile
neutrinos are needed to account for these anomalies.  The possibility thus
remains open that the third could be a keV-scale candidate for dark matter
\cite{Dodelson:1993je,Shi:1998km,Kusenko:2009up,Abazajian:2012ys}.
For the requirements placed by experiment on such a state, see Refs.\
\cite{Boyarsky:2009ix,Merle:2013gea}.  Sterile neutrinos in \esix~have been
considered some time ago (see, e.g., \cite{Rosner:1985hx,Mohapatra:1986aw,%
Nandi:1985uh,Mohapatra:1986bd,London:1986up,Hewett:1988xc,Cvetic:1992ct,%
Ma:1995xk,Chacko:1999aj}), but we discuss them now in the context of present
data (see also \cite{Frank:2004vg,Frank:2005rb,Dev:2012bd}).  A ``minimal
extended seesaw model'' \cite{Zhang:2011vh} has one light sterile neutrino,
rather than three, coexisting with three active neutrinos and their heavy
right-handed counterparts.

Fermion masses in \esix~were analyzed in Ref.\ \cite{Rosner:1985hx}.
(See Ref.\ \cite{Hewett:1988xc} for a review.)  The consequences
were examined of assuming that all masses of fermions in a 27-plet were due to
Higgs bosons in a 27$^*$-plet:  $27 \otimes 27 = 27^* + \ldots$.  A key
shortcoming of this analysis was the lack of a source for large Majorana masses 
of right-handed neutrinos.  Solutions proposed to this problem included
introduction of discrete symmetries, higher-dimension operators, and additional
fermions.  For an overview, including extensive references, see
\cite{Frank:2004vg,Frank:2005rb}.  Our approach is closest to that involving
higher-dimension operators.  E. Ma \cite{Ma:1995xk} showed that a large
Majorana mass for right-handed neutrinos was permitted by a specific scheme of
\esix~breaking which received much subsequent attention from S. F. King and
collaborators \cite{Kinge6}.  (For a recent
reference, see \cite{Callaghan:2012rv}.)  In the present article we update
the analysis of Ref.\ \cite{Rosner:1985hx} and discuss some possibilities for
accommodating the recent suggestions of sterile neutrinos.  We neglect mixing
among fermion families, leaving that topic for further study.  (For one
review, see Ref.\ \cite{King:2014nza}.)

We review some properties of \esix~multiplets, their decomposition into SO(10)
and SU(5) representations, and mass matrix construction in Sec.\ II.  An
\esix-invariant mass matrix for neutrinos is constructed with $27^*$-plet
Higgs fields coupling to $27 \otimes 27$, and its shortcomings pointed out, in
Sec.\ III.  The addition of a large Majorana mass for right-handed neutrinos,
permitted by a specific mode of \esix~breaking, leads to a mass matrix with the
potential to describe conventional very light neutrinos and light sterile
neutrinos mixed with them with arbitrary strength (Sec.\ IV).  The entries of
the neutral lepton mass matrix are compared with corresponding ones for
charged leptons and quarks in Sec.\ V.  The relevance of this scheme to present
results for short-baseline neutrino oscillation experiments is noted in Sec.\
VI.  The presence of three families of quarks and leptons necessitates separate
27-plets for each of them, entailing three states $n$.  Fits to short-baseline
neutrino oscillations \cite{sbl2,sbl3,Kopp:2013vaa} require at least two sterile
neutrinos, leaving the third as a potential candidate for dark matter.  Some
aspects of this identification are discussed in Sec.\ VII.  We conclude in
Sec.\ VIII.
\bigskip

\centerline{\bf II.  REVIEW OF BASICS}
\bigskip

\leftline{\bf A.  \esix~decomposition}
\bigskip

The decomposition of \esix~into SO(10) and SU(5) representations leads to two
U(1) subgroups:  \esix~$\to$ SO(10) $\otimes$ U(1)$_\psi$ and SO(10) $\to$
SU(5) $\otimes$ U(1)$_\chi$ \cite{Langacker:1984dc,London:1986dk}.  The charges
of these two U(1)s are denoted by $Q_\psi$ and $Q_\chi$, and are listed in
Table \ref{tab:list} for left-handed neutral leptons in the first ($e$) family.
In what follows we shall refer exclusively to left-handed states, with
right-handed states related to them by a CP transformation.  The subscript $E$
refers to an exotic vectorlike doublet $(\nu_E,E^-)$ belonging to a
10-dimensional representation of SO(10).  A distinction is made between
neutrinos $\nu$ and their charge-conjugates $N^c$. The charge $Q_N$, defined by
\beq \label{eqn:qndef}
2 \sqrt{10}~Q_N = \frac{5}{4}(2 \sqrt{6}~Q_\psi)
 - \frac{1}{4}(2 \sqrt{10}~Q_\chi)~,
\eeq
is defined for use in Sec.\ IV.  

% This is Table I
\begin{table}
\caption{Neutral leptons in \esix, their SO(10) and SU(5) representations,
and their U(1) charges.  For completeness we also display U(1) charges for
the (16,10) of (SO(10),SU(5)), although it contains no neutral member.
\label{tab:list}}
\begin{center}
\begin{tabular}{c c c c} \hline \hline
   27 member   & $2\sqrt{6}~Q_\psi$ & $2\sqrt{10}~Q_\chi$ & $2\sqrt{10}~Q_N$ \\
(SO(10),SU(5)) &                   &                    &          \\ \hline
$\nu_e(16,5^*)$ &        $-1$       &       ~~3          &       $-2$      \\
~~~(16,10)      &         $-1$      &        $-1$        &       $-1$      \\
$N^c_e(16,1)$   &        $-1$       &       $-5$         &       ~~0       \\
$\nu_E(10,5^*)$ &        ~~2        &       $-2$         &       ~~3       \\
$N^c_E(10,5)$   &        ~~2        &       ~~2          &       ~~2       \\
    $n(1,1)$    &        $-4$       &       ~~0          &       $-5$      \\
\hline \hline
\end{tabular}
\end{center}
\end{table}

Fermion masses \esix~27-plets arise from Higgs bosons transforming as the
product
\beq \label{eqn:tf}
27 \times 27 = 27^* + 351 + 351^*~.
\eeq
Early superstring-inspired models \cite{Witten:1985xc,Candelas:1985en,%
Breit:1985ud,Dine:1985vv} assumed the dominant contributions to come from the
27$^*$-plet, which was the case explored in Ref.\ \cite{Rosner:1985hx}.  We
review that analysis in Sec.\ III and expand it to allow for heavy
right-handed neutrinos in Sec.\ IV.
\bigskip

\leftline{\bf B.  Mass matrices}
\bigskip

In mass matrices involving fermions $f_i$, pairs of left-handed states,
transforming as (1/2,0) under the SU(2) $\otimes$ SU(2) of the Lorentz group,
and pairs of right-handed states, transforming as (0,1/2), must be coupled
to form a Lorentz invariant (0,0).  In a two-component notation, especially
useful in the treatment of neutral particles, 
\beq \label{eqn:2comp}
- {\cal L}_{ij} = \frac{m_{ij}}{2}\epsilon^{\alpha \beta}[(f^c_{iL})_\alpha
(f_{jL})_\beta + (f_{iL})_\alpha (f^c_{jL})_\beta] + (L \to R)~,
\eeq
where $\alpha,\beta = 1,2; \epsilon^{12} = - \epsilon^{21} = 1; \epsilon^{11}
= \epsilon^{22} = 0$.
Consider, for example, $u$ quarks, which are represented by a single field in
each \esix~multiplet.  In a basis described by the fields $u_{aL}=(u_L,u^c_L)$,
the mass term (\ref{eqn:2comp}) then takes the form
\beq \label{eqn:nondiag}
-{\cal L}_m = \frac{1}{2}\epsilon^{\alpha \beta}M_{ab}[(\psi_{a \alpha L}
\psi_{b \beta L}) + (L \to R)]~,
\eeq
where
\beq \label{eqn:mab}
M_{ab} = \left[ \begin{array}{c c} ~0~m_u \cr m_u~0 \end{array} \right]~.
\eeq
Charge conservation prevents $M_{ab}$ from having diagonal entries.
Its eigenvectors and eigenvalues, corresponding to a standard Dirac mass for
$u$, are
\beq
\frac{u_{L,R} + u^c_{L,R}}{\sqrt{2}}~:~~{\rm eigenvalue}~+m_u~,~~~
\frac{u_{L,R} - u^c_{L,R}}{\sqrt{2}}~:~~{\rm eigenvalue}~-m_u~
\eeq

For neutral particles, additional terms in the mass matrix become possible.
For example, the mass matrix $M_{ab}$ in the basis $\nu_L, \nu^c_L$ now can
have diagonal entries.  The left-handed charge-conjugate of $\nu_L$ will be
referred to as $N^c_L$ to denote the fact that both Majorana and Dirac masses
are permitted for neutral leptons.

The popular ``seesaw'' mechanism \cite{seesaw} provides an explanation of why
neutrinos are so light.  Let us restrict the discussion to a single family.  In
the basis $(\nu_L,N^c_L)$ the mass matrix is assumed to take the form
\beq \label{eqn:seesaw}
{\cal M} = \left[ \begin{array}{c c} 0~~m \cr m~M \end{array} \right]~.
\eeq
The Dirac mass terms $m$ transform as SU(2)$_L$ doublets, while the
``right-handed neutrino'' Majorana mass term $M$ transforms as an SU(2)$_L$
singlet and hence is not prevented from taking on a very large value.  The
(unnormalized) eigenvectors and eigenvalues of $M$ are, approximately,
\beq \label{eqn:ssee}
\nu_L - (m/M) N^c_L~:~~{\rm eigenvalue} \simeq -m^2/M~;~~~
(m/M)\nu_L + N^c_L~:~~{\rm eigenvalue} \simeq M~.
\eeq

An equivalent description \cite{Weinberg:1979} is to note the possibility in
the standard electroweak model of a dimension-5 operator describing neutrino
mass bilinear in electroweak doublet Higgs fields $H$, generated by a term
$-{\cal L}_m = (HL)^2/M$, where $L$ is a lepton doublet.  Such a term could
arise, for example, if the heavy right-handed neutrino were integrated out.
The generated neutrino mass would then be of the form $m_\nu = m^2/M$, where
$m$ is of the order of a Dirac mass of quarks or charged leptons, and $M$ is
sufficiently large to yield neutrino mass in the sub-eV range.  Another
approach is to use perturbation theory, with ${\cal M} = {\cal M}_0 +
{\cal M}_1$,
\beq
{\cal M}_0 =  \left[ \begin{array}{c c} 0~~0 \cr 0~M \end{array} \right]~,~~
{\cal M}_1 =  \left[ \begin{array}{c c} 0~~m \cr m~0 \end{array} \right]~
\eeq
and the unperturbed eigenvectors and eigenvectors $[1,0]^T$ (eigenvalue 0)
and $[0,1]$ (eigenvalue $M$).  This is the method we shall use to describe
mixing among more than two neutral leptons when some entries in the mass
matrix are much larger than others.
\bigskip

\centerline{\bf III.  E$_{\rm 6}$-INVARIANT COUPLINGS}
\bigskip

Assuming the mass terms for 27-plets of \esix~in Eq.\ (\ref{eqn:tf}) to
transform as members of the $27^*$-plet, we find a mass matrix for neutral
leptons with non-zero entries indicated in Table \ref{tab:r6}.  The
corresponding mass matrix, assuming it is symmetric and real, is
% This is Table II
\begin{table}
\caption{Values of $(2\sqrt{6}~Q_\psi,2\sqrt{10}~Q_\chi)$ in the product
of two neutral lepton 27-plets represented in a Higgs boson $27^*$-plet.
\label{tab:r6}}
\begin{center}
\begin{tabular}{c|c c c c c} \hline \hline
 & $\nu_e$(--1,3) & $N^c_e$(--1,--5) & $\nu_E$(2,--2) & $N^c_E$(2,2)
 & $n$(--4,0) \\ \hline
$\nu_e$(--1,3) & -- & (--2,--2) & -- & $(1,5)$ & -- \\
$N^c_e$(--1,--5) & (--2,--2) & -- & -- & (1,--3) & -- \\
$\nu_E$(2,--2) & -- & -- & -- & (4,0) & (--2,--2) \\
$N^c_E$(2,2) & (1,5) & (1,--3) & (4,0) & -- & (--2,2) \\
$n(-4,0)$ & -- & -- & (--2,--2) & (--2,2) & -- \\ \hline \hline
\end{tabular}
\end{center}
\end{table}
\beq \label{eqn:r6}
{\cal M}_6 = \left[ \begin{array}{c c c c c} 0 & m_{12} & 0 & M_{14} & 0 \cr
 m_{12} & 0 & 0 & m_{24} & 0 \cr 0 & 0 & 0 & M_{34} & m_{35} \cr
 M_{14} & m_{24}& M_{34} & 0 & m_{45} \cr 0 & 0 & m_{35} & m_{45} & 0
\end{array} \right]
\eeq
in the basis of Table II.
Here we have denoted mass terms transforming as weak isodoublets with
small letters and those transforming as weak isosinglets with large letters.
The latter can take on arbitrarily large values without violating weak
SU(2), while the former are restricted to be less than the electroweak
scale.  The subscript 6 on the mass matrix refers to the rank of the
group under which the couplings of $27 \otimes 27$ to a $27^*$-dimensional
Higgs representation are invariant.  Note that the seesaw term $M_{22}$,
which would have given a large Majorana mass to the right-handed neutrino
$N_e$, is absent.  It corresponds to $Q_\psi,Q_\chi$ values not represented
in a $27^*$-plet.

We shall now make two further assumptions about the properties of ${\cal M}_6$
in Eq.\ (\ref{eqn:r6}).  (1) We shall assume that the exotic vector-like
neutrino $\nu_E$ pairs up with its charge conjugate to obtain a large Dirac
mass, i.e., that $M_{34}$ is very large.  The absence up to the $\sim$TeV
scale of exotic weak isosinglet quarks $h$ with charge $Q = - 1/3$
or vector-like charged leptons $E$ with charge $-1$ supports this assumption.
(2) We shall assume that active-sterile mixing involving $\nu_e$ is small,
as supported by tests of weak universality.  This corresponds to taking
$M_{14}$ relatively small despite its $\Delta I = 0$ nature.

We may analyze the eigenvectors and eigenvalues of the matrix ${\cal M}_6$
by using degenerate perturbation theory, expanding around the corresponding
matrix with only $M_{34}$ nonzero.  It is helpful to first diagonalize
${\cal M}_6$ in $M_{34}$ by a rotation about the 3--4 axis; the result is
\beq \label{eqn:r6p}
{\cal M}'_6 = \left[ \begin{array}{c c c c c}
 0 & m_{12} & M_{14}/\s & M_{14}/\s & 0 \cr
 m_{12} & 0 & m_{24}/\s & m_{24}/\s & 0 \cr
 M_{14}/\s & m_{24}/\s & M_{34} & 0 & (m_{35}+m_{45})/\s \cr
 M_{14}/\s & m_{24}/\s & 0 & -M_{34} & (m_{45}-m_{35})/\s \cr
 0 & 0 & (m_{35}+m_{45})/\s & (m_{45}-m_{35})/\s & 0
\end{array} \right]~.
\eeq
In the limit where all masses except $M_{34}$ are neglected, the eigenvectors
of this matrix with nonzero eigenvalues are $[0,0,1,0,0]^T$ (eigenvalue
$M_{34}$) and $[0,0,0,1,0]^T$ (eigenvalue $-M_{34}$), the hallmarks of a
four-component Dirac spinor.  The three states orthogonal to these are
eigenvectors of a $3 \times 3$ submatrix in the orthonormal basis
$([1,0,0,0,0]^T$, $[0,1,0,0,0]^T$, $[0,0,0,0,1]^T) =(\nu_e,N^c_e,n)$.  Applying
second-order degenerate perturbation theory, we find
\beq \label{eqn:s3}
{\cal S}_3 = \left[ \begin{array}{c c c}
0                    & m_{12}               &  -M_{14}m_{35}/M_{34} \cr
m_{12}               & 0                    &  -m_{24}m_{35}/M_{34} \cr
-M_{14}m_{35}/M_{34} & -m_{24}m_{35}/M_{34} & -2m_{35}m_{45}/M_{34}
\end{array} \right]
\eeq

Under our assumptions, the dominant terms in ${\cal S}_3$ are the two
off-diagonal masses $m_{12}$, leading to eigenvalues $\pm m_{12}$ and
a pseudo-Dirac mass for the conventional neutrino.  The remaining
eigenvalue is approximately $-2m_{35}m_{45}/M_{34}$, associated with
a state which is mostly the sterile neutrino $n$.  As has been noted,
this does not provide a satisfactory picture of very light neutrino masses.
\bigskip

\centerline{\bf IV.  ALLOWING FOR A MASSIVE RIGHT-HANDED NEUTRINO}
\bigskip

E. Ma \cite{Ma:1995xk} has pointed out that the state $N^c_{e}$ has
zero charge under a linear combination of $Q_\chi$ and $Q_\psi$.  In our
notation, this is
\beq
Q_N = -\frac{1}{4}Q_\chi + \frac{\sqrt{15}}{4} Q_\psi~,
\eeq
equivalent to Eq.\ (\ref{eqn:qndef}).  Values of $2 \sqrt{10}~Q_N$ for
neutral leptons in a 27-plet of \esix~are listed in the last column of
Table \ref{tab:list}.

The values of $2 \sqrt{10}~Q_N$ corresponding to products of two neutral
27-plet members are shown in Table \ref{tab:r5}.  The zero value corresponding
to the 22 entry of Table \ref{tab:list} allows for a higher-dimension operator
which breaks U(1)$_\chi$ and U(1)$_\psi$ but preserves their linear
combination $Q_N$.  The corresponding mass matrix ${\cal M}_5$ (with the
subscript denoting the rank of the group under which couplings to Higgs
fields are invariant) is
\beq \label{eqn:r5}
{\cal M}_5 = \left[ \begin{array}{c c c c c} 0 & m_{12} & 0 & M_{14} & 0 \cr
 m_{12} & M_{22} & 0 & m_{24} & 0 \cr 0 & 0 & 0 & M_{34} & m_{35} \cr
 M_{14} & m_{24}& M_{34} & 0 & m_{45} \cr 0 & 0 & m_{35} & m_{45} & 0
\end{array} \right]~.
\eeq
After diagonalization with respect to $M_{34}$, this becomes 
\beq \label{eqn:r5p}
{\cal M}'_5 = \left[ \begin{array}{c c c c c}
 0 & m_{12} & M_{14}/\s & M_{14}/\s & 0 \cr
 m_{12} & M_{22} & m_{24}/\s & m_{24}/\s & 0 \cr
 M_{14}/\s & m_{24}/\s & M_{34} & 0 & (m_{35}+m_{45})/\s \cr
 M_{14}/\s & m_{24}/\s & 0 & -M_{34} & (m_{45}-m_{35})/\s \cr
 0 & 0 & (m_{35}+m_{45})/\s & (m_{45}-m_{35})/\s & 0
\end{array} \right]~.
\eeq
The eigenvectors corresponding to the large eigenvalues, about which we
perturb, are $[0,1,0,0,0]^T$, $[0,0,1,0,0]^T$, and $[0,0,0,1,0]^T$, while
we are concerned with the $2 \times 2$ submatrix ${\cal S}_2$ in the basis
spanned by the eigenvectors $[1,0,0,0,0]^T$ and $[0,0,0,0,1]^T$.  Applying
second-order perturbation theory, we find
\beq \label{eqn:s2}
{\cal S}_2 = \left[ \begin{array}{c c}
   -m_{12}^2/M_{22}    & -M_{14}m_{35}/M_{34} \cr
  -M_{14}m_{35}/M_{34} & -2m_{35} m_{45}/M_{34} \end{array} \right]~.
\eeq

% This is Table III
\begin{table}
\caption{Values of $2\sqrt{10}~Q_N$ in the product
of two neutral lepton 27-plets represented in a Higgs boson $27^*$-plet.
\label{tab:r5}}
\begin{center}
\begin{tabular}{c|c c c c c} \hline \hline
 & $\nu_e$(--2) & $N^c_e$(0) & $\nu_E$(3) & $N^c_E$(2) & $n$(--5) \\
 \hline
$\nu_e$(--2) & -- & --2 & -- & 0 & -- \\
$N^c_e$(0) & --2 & 0 & -- & 2 & -- \\
$\nu_E$(3) & -- & -- & -- & 5 & --2 \\
$N^c_E$(2) & 0 & 2 & 5 & -- & --3 \\
$n$(--5) & -- & -- & --2 & --3 & -- \\ \hline \hline
\end{tabular}
\end{center}
\end{table}

The matrix ${\cal S}_2$ describes the mixing of a conventional neutrino $\nu$
with a sterile neutrino $n$.  The entries are independent of one another,
so arbitrary mixings are possible.  However, an additional constraint is
that present hints of sterile neutrinos place their masses above those
of the three conventional neutrinos \cite{sbl2,sbl3,Kopp:2013vaa}.  So we look
for solutions with small mixing but $m_n > m_\nu$.

If we denote
\beq
\nu = \left[ \begin{array}{c} \cos\theta \cr \sin\theta \end{array} \right]~,~~
 \left[ \begin{array}{c} -\sin\theta \cr \cos\theta \end{array} \right]~,
\eeq
and $t \equiv \tan \theta$, we look for the small-$t$ solution of the
quadratic equation
\beq
t^2 + \left( \frac{m_{12}^2 M_{34}}{M_{14}m_{35}M_{22}}
 - \frac{2 m_{45}}{M_{14}} \right) t - 1 = 0~,
\eeq
in which neglecting the quadratic term gives
\beq \label{eqn:lint}
t \simeq \left( \frac{m_{12}^2 M_{34}}{M_{14}m_{35}M_{22}} 
 - \frac{2 m_{45}}{M_{14}} \right)^{-1}~.
\eeq
Barring accidental cancellation of the two terms, either $|m_{12}^2 M_{34}/
(M_{14}m_{35}M_{22})|$ or $|2 m_{45}/M_{14}|$ must be large.  If $\theta$ is
small, the neutrino mass must be approximately its seesaw value
$m_\nu \simeq - m_{12}^2/M_{22}$.  In order that $m_n > m_\nu$ one
must then have $|2 m_{35}m_{45}M_{22}/(M_{34}m_{12}^2)| > 1$.  But this says
that the term $m_{12}^2 M_{34}/(M_{14} m_{35} M_{22})$ cannot be large
unless $M_{14} \ll m_{45}$.  Thus one can get $m_n > m_\nu$ with small
mixing if
\beq
\left | \frac{m_{35}m_{45}M_{22}}{M_{34}m_{12}^2} \right| > 1~,~~
\frac{m_{45}}{M_{14}} \gg 1~,
\eeq
with no accidental cancellation between the terms on the right-hand side of
Eq.\ (\ref{eqn:lint}).  One cannot take $m_{35} = 0$ if one wants a nonzero
sterile neutrino mass.  The choice of small $M_{14}$ is demanded for
self-consistency of the scheme, but remains an issue of fine-tuning.  It can
be forbidden in lowest order by assigning a $Z_2$ quantum number of $-1$ for
SO(10) 16-plets and $+1$ for SO(10) 10-plets and singlets.  In this manner both
$M_{14}$ and $m_{24}$ are forbidden in lowest order but can be generated by
small $Z_2$-violating vacuum expectation values of SO(10) 16-plet Higgs bosons.

If the terms in Eq.\ (\ref{eqn:lint}) do not destructively interfere with
one another, one or the other will dominate, so that
\beq
t = {\rm min} \left( \frac{M_{14}m_{35}M_{22}}{m_{12}^2M_{34}}, \frac{M_{14}}
{2m_{45}} \right)~.
\eeq
But the first term is larger than the second if $m_n > m_\nu$, so
$t \simeq M_{14}/(2m_{45})$.

The \esix~scheme should be contrasted with a ``minimal extended seesaw''
model \cite{Zhang:2011vh}.  That scheme introduces one sterile neutrino,
rather than three, leading to a $7 \times 7$ mixing matrix when taking
account of three active neutrinos and their right-handed counterparts.
Under some assumptions it can generate either an eV-scale sterile neutrino
to account for short-baseline anomalies, or a keV-scale neutrino as a warm
dark matter candidate.
\bigskip

\centerline{\bf V.  RELATION TO CHARGED LEPTON AND QUARK MASSES}
\bigskip

In grand unification schemes, couplings of Higgs bosons to leptons are
often related to their couplings to quarks at the unification scale.  One must
then apply the renormalization group to evaluate the couplings at accessible
energies.  Such is the case in \esix.  At the unification scale, we shall
see that the neutral lepton parameters $m_{12}$ and $m_{35}$ are related to one
in which quarks of charge 2/3 acquire masses, while the parameters $m_{24}$,
$m_{45}$, $M_{14}$, and $M_{34}$ are related to ones relevant for masses of
quarks of charge --1/3 and charged leptons (both ordinary and exotic).
\bigskip

\leftline{\bf A.  Up-type quarks}
\bigskip

The values of various U(1) quantum numbers for $u$ quarks and their
charge-conjugates are shown in Table \ref{tab:u}, along with charges
of their neutral bilinears.  Referring to Tables II and III, one sees
that these charges correspond to those of the 12, 21, 35, and 53 entries
in the $5 \times 5$ neutral lepton mass matrix:
\beq
(-2,-2,-2) \sim m_{12}, m_{35}~.
\eeq
Thus, the Dirac masses of the electron neutrino and the $u$ quark are related
to one another.  If this relation also holds for the second and third families,
one estimates that the Dirac mass of the heaviest neutrino should be $m_{12}
\simeq (m_\tau/m_b)m_t \simeq 70$ GeV (taking account of renormalization-group
running).  Assuming neutrino masses $m_3 \gg m_2 \gg m_1$, one would expect
from $\Delta m^2_{32} \simeq 2.5 \times 10^{-3}$ eV$^2$ that $m_3 \simeq 5
\times 10^{-2}$ eV and hence a seesaw scale (at least for the heaviest
neutrino) of $M_{22} \simeq 10^{14}$ GeV.

% This is Table IV
\begin{table}
\caption{Charges $2 \sqrt{6}~Q_\psi,2\sqrt{10}~Q_\chi$, and $2\sqrt{10}~Q_N$
of left-handed up-type quarks and antiquarks, and their neutral bilinears.
\label{tab:u}}
\begin{center}
\begin{tabular}{c c c} \hline \hline
                   & $u(-1,-1,-1)$ & $u^c(-1,-1,-1)$ \\ \hline
  $u(-1,-1,-1)$    &       --      &    $(-2,-2,-2)$    \\
$u^c(-1,-1,-1)$ & $(-2,-2,-2)$  &         --         \\ \hline \hline
\end{tabular}
\end{center}
\end{table}
\bigskip
%\newpage

\leftline{\bf Down-type and $h$ quarks}
\bigskip

The left-handed $d,s,b$ quarks quarks are weak isodoublets, while their
charge-conjugates are weak isosinglets.  The left-handed $h$-type quarks and
their charge conjugates are both weak isosinglets.  In Table \ref{tab:d} we
summarize various charges of states and bilinears.  These bilinears have the
same charges as the following entries in the neutral lepton mass matrix:
\bea \label{eqn:assoc}
(-2,2,-3) & \sim & m_{45}~, \nonumber \\
 (1,5,0)  & \sim & M_{14}~, \nonumber \\
 (1,-3,2) & \sim & m_{24}~, \nonumber \\
 (4,0,5)  & \sim & M_{34}~.
\eea
In particular, in the absence of mixing of ordinary and exotic quarks, $m_{45}$
is related to the Dirac mass of ordinary down-type quarks, while $M_{34}$ is
related to the Dirac mass ($> {\cal O}(1)$ TeV) of $h$-type quarks.  Weak
universality suggests that the mixing of weak isodoublet left-handed $d$-type
quarks with weak left-handed isosinglet $h$-type quarks is small, and hence in
grand unification schemes that $m_{24} \ll m_{45}$.  Since $M_{14}$ is related
to a quantity which mixes weak-isosinglet ordinary quarks with weak isosinglet
exotic ones, there seems to be less of a constraint on that matrix element
coming from quarks of charge --1/3 and their antiquarks.  For some discussions
of $h$-quark properties and searches, see Refs.\ \cite{Rosner:2000rd,%
Bjorken:2002vt,Andre:2003wc}.

% This is Table V
\begin{table}
\caption{Charges $2 \sqrt{6}~Q_\psi,2\sqrt{10}~Q_\chi$, and $2\sqrt{10}~Q_N$
of left-handed $d$- and $h$-type quarks and their charge conjugates, along
with charges of their neutral bilinears.
\label{tab:d}}
\begin{center}
\begin{tabular}{c c c c c} \hline \hline
 & $d^c(-1,3,-2)$ & $d(-1,-1,-1)$ & $h^c(2,-2,3)$ & $h(2,2,2)$ \\ \hline
 $d^c(-1,3,-2)$  &      --     & $(-2,2,-3)$ &     --    & $(1,5,0)$ \\
    $d(-1,-1,-1)$   & $(-2,2,-3)$ &      --     & $(1,-3,2)$ &     --    \\
  $h^c(2,-2,3)$  &     --      &  $(1,-3,2)$ &     --    & $(4,0,5)$ \\
  $h(2,2,2)$   &  $(1,5,0)$  &      --     & $(4,0,5)$ &-- \\ \hline \hline
\end{tabular}
\end{center}
\end{table}
\bigskip

\leftline{\bf Charged leptons}
\bigskip

The pattern of mixing of charged leptons resembles that
associated with quarks of charge --1/3 and their antiquarks.  The U(1)
charges of states and bilinears are summarized in Table \ref{tab:e}.
The same association of neutral lepton mass matrix entries with U(1)
charges exhibited in Eq.\ (\ref{eqn:assoc}) holds here as well.  In
the absence of ordinary-exotic mixing, $m_{45}$ is associated with an
ordinary charged-lepton Dirac mass, and $M_{34}$ is associated with an
exotic (``$E$-type'') lepton Dirac mass.  The analog of $m_{24}$ mixes
a weak isosinglet ordinary-lepton mass with a weak isodoublet exotic-lepton
mass and thus must be much smaller than the analog of $m_{45}$, while the
analog of $M_{14}$ mixes weak isodoublets with one another and thus is not
strongly constrained.

% This is Table VI
\begin{table}
\caption{Charges $2 \sqrt{6}~Q_\psi,~2\sqrt{10}~Q_\chi$, and $2\sqrt{10}~Q_N$
of left-handed charged leptons, along with charges of their neutral bilinears.
\label{tab:e}}
\begin{center}
\begin{tabular}{c c c c c} \hline \hline
 & $e^-(-1,3,-2)$ & $e^+(-1,-1,-1)$ & $E^-(2,-2,3)$ & $E^+(2,2,2)$ \\ \hline
$e^-(-1,3,-2)$  &      --     & $(-2,2,-3)$ &     --    & $(1,5,0)$ \\
$e^+(-1,-1,-1)$   & $(-2,2,-3)$ &      --     & $(1,-3,2)$ &     --    \\
$E^-(2,-2,3)$  &     --      &  $(1,-3,2)$ &     --    & $(4,0,5)$ \\
$E^+(2,2,2)$   &  $(1,5,0)$  &      --     & $(4,0,5)$ &-- \\ \hline \hline
\end{tabular}
\end{center}
\end{table}

\bigskip

\centerline{\bf VI.  RELATION TO SHORT-BASELINE OSCILLATION EXPERIMENTS}
\bigskip

So far we have considered mixing of $\nu$ and $n$ within a single family,
finding enough freedom in \esix~to write a generic $2 \times 2$ matrix with
arbitrary terms as long as we permit a large seesaw mass $M_{22}$.  The
most general mass matrix for three families will then be $6 \times 6$.  As
has been pointed out in Ref.\ \cite{sbl2}, it is sufficient to neglect the
mass differences among light neutrinos when discussing oscillations sensitive
to squared mass differences in the eV$^2$ range.  Consequently, we may
speak of squared mass differences $\Delta m_{41}$, $\Delta m_{51}^2$, and
$\Delta m_{61}^2$ (in increasing order), and mixing matrices $U_{\alpha i}$,
where $\alpha = (e,\mu,\tau)$ and $i = (4,5,6)$.  There will also be
CP-violating phases when there are at least two sterile neutrinos.  These can
be useful when accounting for differences between neutrino and antineutrino
oscillations.

In Ref.\ \cite{sbl2} a model with three active and $N$ sterile neutrinos will
be referred to as a $3 + N$ model.  In fits to a $3+1$ model, neutrino and
antineutrino data favor very different oscillation parameters, as do appearance
and disappearance data.  The probability of compatibility among all data is
rated as 0.043\%.  This probability rises to 13\% in a $3+2$ model and 90\% in
a $3+3$ model.  However, in $3+2$ and $3+3$ models the compatibility of
appearance and disappearance data is still only about 0.008\%.  This is mainly
due to a poor fit to the low-energy signal of electron neutrino and
antineutrino appearance data in the MiniBooNE experiment
\cite{AguilarArevalo:2007it}.  It is still not settled whether those data are
due to electrons (positrons for antineutrinos) or to photons, e.g., from an
anomalous coherent $Z-\gamma$ interaction with the target nucleus
\cite{Harvey:2007rd}.  The main improvement associated with the $3 + 3$ model
appears to be increased compatibility (53\%) of neutrino vs.\ antineutrino
data, compared with 5.3\% for the $3+2$ model.

The fits assume $U_{\tau i} = 0$, allowing for maximal values of $|U_{ei}|$
and $|U_{\mu i}|$.  Typical values of these parameters in both $3+2$ and
$3+3$ models are about $0.15 \pm 0.05$.  In both models $\Delta m_{41}^2$
is about 0.9 eV$^2$ and $\Delta m_{51}^2$ is about 17 eV$^2$, while an
additional state with $\Delta m_{61}^2 = 22$ eV$^2$ is present in the
$3 + 3$ model.  In the $3+2$ model, there is an additional allowed region
with $\Delta m_{51}^2 \simeq 0.9$ eV$^2$.

The fits of Ref.\ \cite{Kopp:2013vaa} also favor more than one sterile
neutrino, with preference for a scheme with two extra neutrinos having
$\Delta m_{14}^2 = -0.87$ eV$^2$ and $\Delta m_{15}^2 = 0.47$ eV$^2$.
Mixing parameters are in the same rough range as those in Ref.\ \cite{sbl2}.
Again, attention is called to the tension between neutrino appearance and
disappearance data.

A cautionary note has been sounded \cite{Hayes:2013wra} with regard to a
claimed 6\% deficit in the flux of reactor neutrinos \cite{Mueller:2011nm,%
Huber:2011wv}.  It has been pointed out that calculations of these fluxes did
not take account of the uncertainty associated with the 30\% of the flux that
arises from forbidden decays.

How does this relate to the $3 + 3$ scenario predicted by \esix?  The
freedom we have to describe a single family is an encouraging sign
that similar freedom might exist in the three-family case where mixings
are governed by a $6 \times 6$ light-neutral-lepton mass matrix.  However,
without a basic understanding of the mixings of $\nu_e,\nu_\mu,\nu_\tau$,
it may be premature to attempt such a description.  The question also is
not yet settled whether a third neutrino is needed to help describe
short-baseline oscillation phenomena, or is available as a dark matter
candidate. 

We established that typical mixings between light conventional neutrinos
and sterile ones are of order $M_{14}/(2 m_{45})$.  In order to accommodate
suggestions of $|U_{ei}|$ and $|U_{\mu i}|$ of order $0.15 \pm 0.05$,
this ratio of mass parameters must be of the same order.  As the analogue
of $m_{45}$ describes masses of down-type quarks or charged leptons, $M_{14}$
must be a small but non-negligible fraction of this quantity.  This feature is
perhaps the most finely-tuned aspect of the present framework.  As mentioned,
assignment of a $Z_2$ quantum number of $-1$ to SO(10) 16-plets and $+1$
to 10-plets and singlets is one way to achieve this suppression.
\bigskip

\centerline{\bf VII.  THIRD STERILE NEUTRINO AS A DARK MATTER CANDIDATE}
\bigskip

Light sterile neutrinos with masses in the keV range have been proposed to
account for some or all of the dark matter in the Universe
\cite{Dodelson:1993je,Shi:1998km,Kusenko:2009up,Abazajian:2012ys}.  (There may
be more than one species \cite{Rosner:2005ec}.)  Two recent claims for
keV-scale dark matter are based on the observation of an X-ray line near 3.5
keV \cite{Bulbul:2014sua,Boyarsky:2014jta} which could arise from the decay of
a 7 keV neutral lepton to a photon and a much lighter neutral lepton.  (The
absence of such a line in the Milky Way is a source of caution
\cite{Riemer-Sorensen:2014yda}.)

Constraints arising from taking a keV-scale neutrino to be a source of warm
dark matter, and proposals for its production, have been reviewed in Ref.\
\cite{Merle:2013gea}. Recent proposals involving a 7 keV neutrino include those
in Refs.\ \cite{Ishida:2014dlp,Abazajian:2014gza,Bezrukov:2014nza,%
Robinson:2014bma,Chakraborty:2014tma}.  Values of the mixing parameter between
the sterile neutrino and a light one are typically somewhat below $\sin^2 2
\theta = {\cal O} (10^{-10})$, which is easily accommodated in an \esix~model.
In comparison with models (see, e.g., \cite{Merle:2013gea}) in which a
keV-scale neutrino is taken to be one of the three right-handed neutrinos, such
a scenario affords greater freedom in choice of parameters.

Some remarks on the special aspects of an \esix~framework for keV-scale dark
matter are in order.  The Higgs vacuum expectation values we have introduced
correspond to five neutral complex scalar bosons belonging
to the $27^*$ representation of \esix.  The masses of these bosons are free
parameters.  The standard model Higgs boson {\it happens} to have a mass
which is close to $1/\sqrt{2}$ of its vacuum expectation value, but there is
no reason for this to be true in general.  (In fact, two of the five neutral
Higgs bosons are just those of the Minimal Supersymmetric Standard Model or
any left-right symmetric model including SO(10).)  But exchanges of these
bosons can give rise to exotic processes producing the state $n$: e.g.,
\beq
d_l + h^c_L \to n_L + N^c_{EL}~~;~~~e^- + E^+_L \to n_L + N^c_{EL}
\eeq
Furthermore, if a TeV-scale $Z_N$ is produced in the early universe,
it would have an appreciable branching ratio for decay into $n n^c$,
according to the $Q_N$ quantum numbers listed in Table I.  Thus $n$ are
candidates for early overproduction unless their abundance is diluted
by subsequent entropy production \cite{Scherrer:1985zt}.

\bigskip

\centerline{\bf VIII.  CONCLUSIONS}
\bigskip

The present paper is not meant to be a definitive model for sterile neutrinos
in \esix, but rather an answer to the question: ``What does it take for \esix~%
to be a satisfactory framework for treating sterile neutrinos?''  The following
conditions have been found sufficient (though others may well exist):

\begin{itemize}

\item The standard seesaw mechanism accounts for all three (light) active
neutrino masses, entailing three very heavy right-handed neutrinos
which are left-handed SU(2) singlets and right-handed SU(2) doublets.

\item Fermion masses arise from the lowest-dimension (27$^*$) Higgs
representation in \esix, aside from a mechanism permitting right-handed
neutrinos to acquire large Majorana masses.

\item This is achieved by breaking \esix~down to SU(5) $\times$ U(1)$_N$, where
U(1)$_N$ is a symmetry under which right-handed neutrinos are neutral.  Q$_N$
is the corresponding charge.

\item Exotic isodoublet leptons $\nu_E,E$ and isosinglet quarks $h$ should
acquire large Dirac masses and mix weakly with lighter states.

\item A term $M_{14}$ in the $5 \times 5$ neutrino mass matrix is taken to be
small despite carrying zero weak (left-handed) isospin.  This fine-tuning,
possibly achieved via a $Z_2$ symmetry, seems needed to describe the observed
spectrum.

\end{itemize}

The grand unified group \esix~provides candidates for three light sterile
neutrinos.  At least two of these seem useful to
account for present-day anomalies in short-baseline neutrino oscillation
experiments.  (It goes without saying that these are urgently in need of
confirmation, as none rises to the level of a ``discovery.'')  A third may
improve the description of these anomalies, or could be available as a
candidate for dark matter, such as suggested by recent X-ray observations.
The extended Higgs and gauge structure of \esix~permits new
mechanisms for production of these candidates in the early universe.

Conclusive evidence for three sterile neutrinos, if interpreted within the
framework of \esix, would entail also isovector charged leptons and isosinglet
quarks $h$ with charge --1/3.  These would then be prime targets for
higher-energy searches at the CERN Large Hadron Collider.
\newpage
% \bigskip

\centerline{\bf ACKNOWLEDGMENTS}
\bigskip

I thank Janet Conrad, P. S. Bhupal Dev, Mariana Frank, and Rabi Mohapatra for
pointing me to some helpful literature, and Kevork Abazajian, Joshua Frieman,
Richard Hill, Lauren Hsu, Hitoshi Murayama, and Robert Shrock for useful
discussions.  This work was supported in part by the United States Department
of Energy through Grant No.\ DE FG02 13ER41958.

\end{document}